\documentclass[3p,times]{elsarticle}

\usepackage{url}
\usepackage[utf8]{inputenc}
\usepackage{amssymb}
\usepackage[figuresright]{rotating}
\usepackage{multicol}
\begin{document}
\begin{frontmatter}
\title{Sipping Science in a Café}

\author[en,csdc,cs]{Franco Bagnoli}
\ead{franco.bagnoli@unifi.it}
\author[csdc,cs]{Giovanna Pacini}
\ead{giovanna.pacini@complexworld.net}

\address[de]{Dept Energy and CSDC, University of Florence
via S. Marta 3, 50139 Firenze,Italy
2. Also  INFN, sez. Firenze}

\address[csdc] {CSDC, University of Firenze, via G. Sansone 1, 50019 Sesto Fiorentino, Italy}   \address[cs]{Associazione Culturale Caffè-Scienza, c/o CNR-ISC, Via Madonna del Piano 10, 50019 Sesto Fiorentino, Italy}

\begin{abstract}
We present here the European project SciCafé - networking of science cafés in Europe and neighboring countries, and the contributions of the CSDC-CaffèScienza partner in Florence, Itay.
\end{abstract}


\end{frontmatter}



\begin{figure}[t]
\begin{center}
 \begin{tabular}{cc}
\includegraphics[height=0.25\textheight]{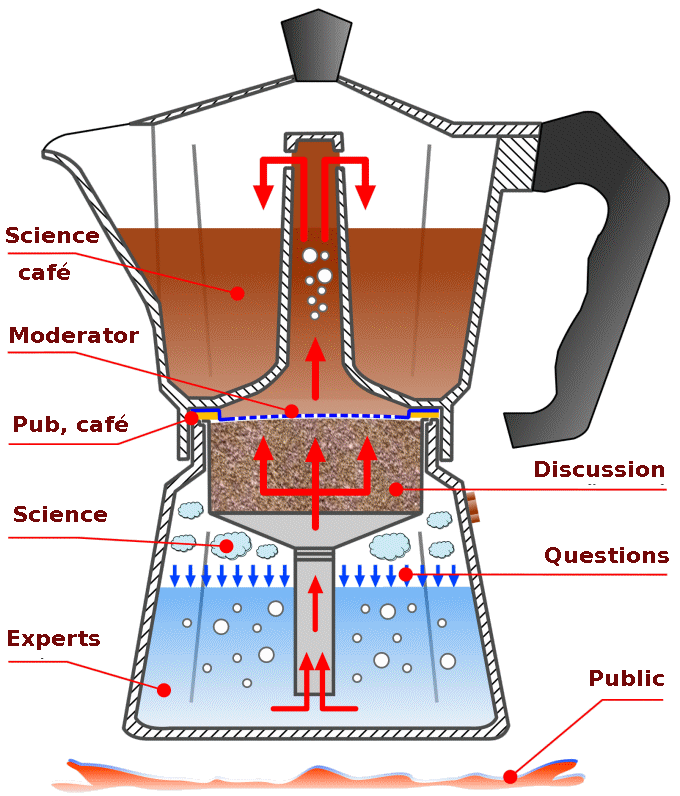} & \includegraphics[width=0.65\columnwidth]{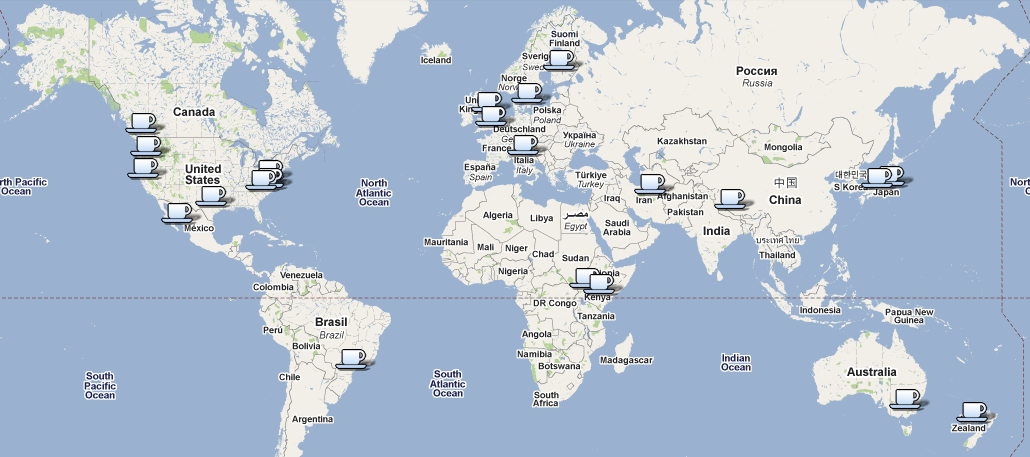}
 \end{tabular}
\end{center}
\caption{\label{moka} (left) A joke about the double meaning (in Italian) of the word ``caffè'': it means a café (place) and a coffee, so a science café (caffè-scienza) may indicate a discussion or a special coffee brand. The Italian name for the coffee machine is moka. (right) The location of the contributors to the science café web book.} 
\end{figure}

Imagine yourself into a pub or a café, drinking a beer or a tea. You are relaxed and almost happy. You may ask yourself: what am I missing now? Science! of course.
A science café is a discussion about some topic in science and technology as scientists do. This does not mean drawing formulas on napkins, but discussing with experts all on the same ground, where the attendees, and not the experts, are at home. In other words, a science café is a conference “upside down”: hosted in a pub or in a café (but never in a conference hall), it generally starts with a short introduction by experts that present themselves and the subject of the discussion, after which the microphone is offered to the public and the rest of the event is driven by questions.
The machinery of a science café is illustrated in the  Fig~\ref{moka}. 

In Florence, Italy, this activity is carried out by the non-profit association ``Caffè-Scienza''~\cite{caffescienza}, formed by  academic and CNR researchers, but also many ``ordinary'' people.
Our association organizes the traditional science cafés (caffè-scienza) monthly and other initiatives such as the junior science cafés (with high school students). As in other science cafés, our public is mainly composed by middle-age, highly educated people. Since 2005, an average of 60 people attends our 6-7 caffè-scienza each year, with some success (last year, our ex-president Paolo Politi won the Italian Physics Society award for science communication). But recently a new wave started.
 
In 2007, many people from several science cafés in Europe met in Ajaccio (Corse) for the $10^{\mbox{th}}$ anniversaire of the network \emph{Bars et Cafés des sciences}. In this occasion it was proposed to apply for a suitable UE grant for the constitution of an European network of science cafés. There were already networks among UK, USA and French science cafés, but no European coordination. Moreover, these networks mainly deal with a presentation of the various locations. 
In January 2010, the European Project \emph{Scicafé}~\cite{scicafe} has started.  The project's main target is to create a European network of science cafés. We aim to identify the best practices used in this kind of scientific popularization and participation, promote the birth of new cafés, especially in eastern countries and in Africa; enlarge the audience of our events to young people and other classes. 

The project is essentially in the phase of data collecting, however some of us have started experimenting  new techniques that resulted to be quite interesting, and forced us to learn new skills. we shall illustrate some of these experiments carried out by our association, in collaboration with others. Most of them were suggested by our public through a questionnaire, that is now being administered to an European public through our partners in the consortium. 

\textbf{Cafferenze}. 
Sometimes an interesting theme is not suited for a science café, mainly because it is too technical and the public may have a few questions to ask. So we launched an hybrid between a science café (caffè-scienza) and a conference (conferenza), that we called “cafferenza”. They are quite successful (and hosted in a beautiful library~\cite{oblate}). 
\textbf{Moka and RadioMoka}.
We started a newsletter (Moka, the Italian name for the coffee machine in Fig.~\ref{moka}-left) and a radio transmission (RadioMoka~\cite{novaradio}), this last experience really required quite different skills from those that we studied in the university!
\textbf{Audio and video streaming} We are collaborating with RadioSpin~\cite{radiospin}, one of the web radios of the University of Firenze, and, with our twin association formascienza in Rome, experimenting with the video streaming service offered by one of our partners, DBC-tv~\cite{dbc} and other independent services~\cite{livestream}. Thanks to DBC-tv, now our events can be attended even from Second Life, and actually we have a few people that regularly follows us from Spain, Switzerland and USA.
\textbf{The science café web book}.
It is just a  web site~\cite{webbook}, assembled 
in collaboration with Duncan Dallas from Leeds, who is essentially the ``inventor'' of science cafés in UK.  We felt that we were concentrating too much on the technical aspects, forgetting that a science café is mainly driven by passions and emotions. So we asked people from all the world to send us a few lines about their motivations (see Fig.~\ref{moka}-right), and we discovered that the same simple concept of a science café can indeed be interpreted in very different ways, from a discussion on how to prevent HIV infection in Uganda, to gender differences in Iran, to a specialized Nuclear Science Café in Oregon!
\textbf{Survey on Science Cafè's Public}. 
In November 2009, we sent an email to our mailing list asking to fill out an online survey. The purpose of the survey was to gather real data about the working and perception of our science café, to identify the effective ways for its advancement and growth.
In particular, we sought to know the type of audience that attended the meetings and receive suggestions about new formats to be used in meetings.
In February 2011 we shared the survey with the partners of the European Project in order to highlight the differences and similarities among different countries and approaches.


Science and technology are more and more important in our lives, and we are often asked to choose (or vote) on technical questions. A science discussion is often seen as a popularization event, but it should be rather termed participation. The purpose of the science café is that of demythologizing science communication, bringing it out of the cathedra and into everyday life. The scicafé European project will be hopefully useful for this goal, favoring the birth of new cafés in new locations, favouring discussion through new media, involving new public and networking all such experiences.


We acknowledge useful discussions with your partners in the project SciCafé, in particular with Tommaso Castellani of FormaScienza, Rome, and Duncan Dallas of Leeds, UK.






\begin{thebibliography}{00}



\bibitem{caffescienza} \url{http://www.caffescienza.it}
\bibitem{scicafe} \url{http://www.scicafe.eu}
\bibitem{oblate} \url{http://www.bibliotecadelleoblate.it}
\bibitem{novaradio} \url{http://www.novaradio.info}
\bibitem{radiospin} \url{http://www.radiospin.poloprato.unifi.it}
\bibitem{dbc} \url{http://www.dbc-tv.net}
\bibitem{livestream} \url{http://www.livestream.com}
\bibitem{webbook} \url{http://sites.google.com/site/scicafewebbook}

\end{thebibliography}



\begin{multicols}{2}

\end{multicols}
\end{document}